\documentclass[runningheads, colorinlistoftodos]{llncs}
\usepackage[T1]{fontenc}
\usepackage{graphicx}
\usepackage[textsize=scriptsize, disable]{todonotes} %, disable
\setuptodonotes{tickmarkheight=0.2cm}
\usepackage{outlines}

\usepackage{cite} % Allows, i.A., multi-cite (with spaces), 
\usepackage{hyperref} % Clickable references 
\usepackage[nameinlink]{cleveref} % Allows, e.g., capitalization of references at sentence starts
\crefname{figure}{Fig.}{Figs.}
\usepackage{float} % Figures that don't jump around
\usepackage{comment}
\usepackage{enumitem} % RQ enumeration
\usepackage{svg}
\usepackage[normalem]{ulem} % Strikethrough
\makeatletter
\def\newparshape{\parshape\@npshape0{}}
\def\@npshape#1#2#3{\ifx\\#3\expandafter\@@@npshape\else\expandafter\@@npshape\fi
  {#1}{#2}{#3}}
\def\@@npshape#1#2#3#4#5{%
  \ifnum#3>\z@\expandafter\@firstoftwo\else\expandafter\@secondoftwo\fi
  {\expandafter\@@npshape\expandafter{\the\numexpr#1+1\relax}{#2 #4 #5}{\numexpr#3-1\relax}{#4}{#5}}%
  {\@npshape{#1}{#2}}}
\def\@@@npshape#1#2#3{#1 #2 }
\makeatother

\newcommand{\betterwrapfig}[3]{
\newparshape
  {#2}{\dimexpr#1+.05\textwidth+\tabcolsep}{\dimexpr\textwidth-#1-\tabcolsep-.05\textwidth}% Unique to top-left image
  {1}{0pt}{\textwidth}% full-width lines where no image is present
  {0}{0pt}{0}\\% Unique to bottom-right image
\noindent\leavevmode
\llap{%
  \raisebox{\dimexpr-\height+\ht\strutbox}[0pt][0pt]{%    
    \fcolorbox{white}{white}{
    \begin{minipage}{#1-\tabcolsep-\tabcolsep}
    #3
    \end{minipage}}
    }
\hspace*{\tabcolsep} %
} %
}

\usepackage[newfloat, cachedir=./minted, frozencache]{minted}
\usepackage{caption}
\newenvironment{code}{\captionsetup{type=listing, skip=1pt}}{}
\counterwithout{listing}{chapter}
\SetupFloatingEnvironment{listing}{name=Listing}

\usepackage{algorithm}
\usepackage{algpseudocode} % algorithmicx
\algtext*{EndFor}
\algtext*{EndProcedure}
\algtext*{EndIf}

\usepackage{tikz}
\usetikzlibrary {graphs,quotes,positioning,calc,fit,backgrounds} % For Graph inference rules
\tikzstyle{arrow} = [thick,->,>=stealth, line width=1.1pt, line cap=round]
\tikzstyle{entity} = [rectangle, rounded corners, line width=1.1pt, draw, inner sep=1mm,outer sep=0, minimum size=.5cm, anchor=west, align=center]
\tikzstyle{varentity} = [entity, dashed]
\tikzstyle{relation} = [anchor=west, align=center]
\usepackage{relsize}

\newcommand{\quotes}[1]{``#1''}
\newcommand{\squotes}[1]{`#1'}

\newcommand{\iodo}[2][]{\todo[color=red!40,#1]{IMPORTANT: #2}} % Important todo
\newcommand{\lodo}[2][]{} % Less important todo ;)
\newcommand{\LP}[2][]{\todo[color=blue!60!red,#1]{\color{white}LP: #2}} % Comments by Luise
\newcommand{\optnewpage}[0]{\newpage} 
\renewcommand{\optnewpage}{} % Uncomment this line for final polishing

\let\oldoutline\outline
\let\oldendoutline\endoutline
\renewcommand{\outline}{\bgroup\color{blue!60!red}\oldoutline}
\renewcommand{\endoutline}{\oldendoutline\egroup}

\usepackage{fmtcount}
\usepackage{totcount}

\newtotcounter{obj}
\crefformat{obj}{#2O#1#3}
\crefrangeformat{obj}{#3O#1--#2#4)}
\newcommand{\objective}[1]{{\refstepcounter{obj}\label{#1}}\item[\cref*{#1}]}%\autoref*{#1} #2}}

\begin{document}
\title{KRAFT -- A Knowledge-Graph-Based Resource Allocation Framework}
\author{Leon Bein\inst{1}\orcidID{0000-0001-9064-7905} \and
Niels Martin\inst{2}\orcidID{0000-0003-3279-3853} \and
Luise Pufahl\inst{1}\orcidID{0000-0002-5182-2587}} 
\authorrunning{L. Bein, N. Martin, and L. Pufahl}
\institute{School of Computation, Information, and Technology, 
\\ Technical University of Munich, Heilbronn, Germany
\\\email{<firstname.lastname>@tum.de}
\and
Hasselt University, Hasselt, Belgium
\\\email{<firstname.lastname>@uhasselt.be}
}

\newcommand{\anonymize}[0]{
\author{{\color{white}.\\.}Anonymized for Double-blind{\color{white}\\.\\.}} % Empty lines to ensure same space usage as multiple authors 
\authorrunning{Anonymized}
\institute{Department, 
\\ University
\\\email{name@institution.edu}}
}

%\anonymize

% ====== Document Start ========
%
\maketitle              % typeset the header of the contribution
\begin{abstract}

Resource allocation in business process management involves assigning resources to open tasks while considering factors such as individual roles, aptitudes, case-specific characteristics, and regulatory constraints. Current information systems for resource allocation often require extensive manual effort to specify and maintain allocation rules, making them rigid and challenging to adapt. In contrast, fully automated approaches provide limited explainability, making it difficult to understand and justify allocation decisions. Knowledge graphs, which represent real-world entities and their relationships, offer a promising solution by capturing complex dependencies and enabling dynamic, context-aware resource allocation. This paper introduces KRAFT, a novel approach that leverages knowledge graphs and reasoning techniques to support resource allocation decisions. We demonstrate that integrating knowledge graphs into resource allocation software allows for adaptable and transparent decision-making based on an evolving knowledge base.

 %\todo{LB: We might add the universality/holisticness and explainability aspects here}

\end{abstract}
\section{Introduction}\label{sec:intro}

Organizations execute business processes to deliver products or services to their customers. Effective and efficient process execution % of business process activities 
relies on various resources, including staff, machines, vehicles, and software that perform the work of process activities~\cite{dumas2018FundamentalsBusinessProcess}. 
These resources are diverse and have varying capabilities and functionalities. At the same time, they are valuable, often costly, and scarce. 

During process execution, each newly enabled task, i.e., activity instance, requires allocating one or more suitable resources, a decision that is inherently complex~\cite{pufahl2025automatic}. 
For human resources, e.g., factors such as individual expertise, past experience, and team membership, but also their availability, task-specific characteristics, other competing tasks, etc. can influence the speed, outcome, and quality of task execution~\cite{arias2018TaxonomyHumanResource}. 
Additionally, dependencies between preceding and subsequent tasks within the process further complicate resource allocation\cite{xu2013incorporating}. We refer to such factors as \emph{knowledge} that needs to be considered during resource allocation.

While resource allocation can be performed manually, information systems and business process management systems (BPMS) typically provide capabilities to support this complex decision~\cite{dumas2018FundamentalsBusinessProcess,pufahl2025automatic}.
In many cases, process experts then define resource allocation knowledge as extensions to a process model, e.g., by specifying rules and constraints for process activities~\cite{russel2004workflow,cabanillas2015SpecificationAutomatedDesigntime}, formulating allocation problems and solution techniques~\cite{ihde2022framework}, or incorporating additional visual elements~\cite{cabanillas2015ralph}. Furthermore, process experts must define the set of available resources along with their attributes and, in some cases, the organizational structure, such as hierarchical relationships~\cite{zurmuehlen2004OrganizationalManagementWorkflow}. 
This approach often requires \emph{significant manual effort} for specification and raises \emph{maintenance challenges}, as business processes are regularly adapting to changing environments~\cite{reichert2012enabling}, also necessitating changes in resource allocation. 
% \LP{Should we include the point by Leon: certain domain-specfic aspects might not be so easy be specified by the approaches proposed?} 
%Finally, following from and complementing the previous points, 
Approaches have been developed to extract resource allocation knowledge from past process execution data %(using, e.g., machine learning and reinforcement learning) 
and use this knowledge for real-time resource allocations \cite{pufahl2025automatic}. 
However, such approaches are limited in the concepts for which knowledge can be extracted \cite{rubensson2025ConceptualFrameworkResource}. 
Further, \emph{explainability} is necessary for complex process-decision-making systems \cite{dumas2023AIAugmentedBusinessProcess} to build trust in the system's decisions and is missing in current resource allocation support~\cite{pufahl2025automatic}. 
% \NM{When I read this as a reviewer, this suggests that our contribution is only explainability. However, I assume that also the approaches mentioned in the prior sentence still have some limitations (e.g. limited scope in terms of concepts for which knowledge can be extracted)? LB: Added an appropriate sentence}. 
System-provided reasoning allows human decision-makers to make informed decisions or adapt system-made decisions while saving the time to look up the relevant information\cite{tiddi2022KnowledgeGraphsTools}.

Methods from the fields of knowledge graph reasoning and ontology-based business process modeling appear promising to bridge this gap. 
Knowledge graph reasoning aims to derive new knowledge from rich graph representations of the real world \cite{chen2022OverviewKnowledgeGraph}, providing tools that can consider all that knowledge in an explainable way \cite{delong2024NeurosymbolicAIReasoning, tiddi2022KnowledgeGraphsTools}. 
% \NM{Connection with the rest of the paragraph might not be clear as KG was not mentioned before. Shouldn't we introduce KG more explicitly as a potential solution as this is the central building block of KRAFT?}
In turn, ontology-based business process modeling is concerned with extending classical process models with instances of ontologies that represent additional background process knowledge \cite{corea2021OntologyBasedProcessModelling}, resulting in rich graph structures of process knowledge.

\paragraph{Research Objectives}
While existing works, e.g., propose to consult taxonomies of resource criteria~\cite{ouyang2010ModellingComplexResource, arias2018TaxonomyHumanResource} or propose the usage of graph reasoning to recommend next activities \cite{bein2025KnowledgeGraphsKey}, the usage of knowledge graph reasoning in combination with ontological modeling 
has not yet been investigated for resource allocation in business processes. 
To this end, this paper investigates this usage, considering the following objectives:

\begin{itemize}[topsep=-\parskip, itemsep=1mm]      
    \objective{obj:hol} It should support capturing and encoding a broad set of allocation knowledge%
    %
    %\NM{A word seems to be missing here. How do we want to call concepts such as experience, availability,...? In the Weave proposal, I call them RA decision attributes, but I'm fine with another term as well.\\LB: Doesn't allocation knowledge do the job here?}
    %}
    \objective{obj:adap} It should support an easy adaptation of the encoded allocation knowledge %
    %\LB{Could cut mechanisms here as we consider mechanism knowledge to be part of allocation knowledge <=> could also leave to not "spoil the surprise"}
    \objective{obj:expl} It should determine (the most) suitable resources based on that knowledge and comprehensively explain the decisions made%
\end{itemize}  

\noindent To realize these objectives, this paper proposes \emph{KRAFT, a Knowledge-graph-based Resource Allocation Framework for Tasks} in business processes. It allows populating, updating, and reasoning on a knowledge graph that captures resource allocation knowledge for explainable resource allocation decision support. 
%this paper presents the KRAFT framework%, the knowledge-graph-based resource allocation framework.  
Further, we present a prototypical implementation and demonstration of it on a loan application process, provided as a real-world data set in~\cite{vandongen2017bpic}. \lodo{Could condense here even more by integrating the sentence on demo into the structure of the paper OR delineate better, by adding the classical "the remainder of this paper ..."}
In \Cref{sec:background}, we discuss preliminaries and related work on resource allocation and knowledge graphs. Then, in \Cref{sec:main}, KRAFT is presented in detail, and demonstrated using a proof of concept implementation in \Cref{sec:poc}. Finally, we discuss our findings in \Cref{sec:disc} and conclude in \Cref{sec:concl}. 

\begin{comment}
\paragraph{Method \& Structure of the Paper}
\begin{outline}
    \1 We draw from design science research \cite{peffersDesignScienceResearch2007} 
    \1 Having outlined the problem to solve and our objectives above \iodo{Please do!},
        \2 we will discuss preliminaries on resource allocation and knowledge graphs in \cref{sec:background},
        \2 present our framework in detail in \cref{sec:main},
        \2 demonstrate it using a proof of concept implementation in \cref{sec:poc}
    \1 Discussing preliminaries in \cref{sec:background} and relevant existing works in \cref{sec:relwork}
\end{outline}
\end{comment}

\optnewpage
\section{Background \& Related Work} \label{sec:background}
\subsection{Resource Allocation} \label{sec:resall}
% \todo{Maybe just cite fundamentals/weske books? => Nope, surprisingly unuseful here}
% In this paper, we view resource allocation from the lens of business process management (BPM), which is concerned with ...
In the business process management (BPM) field, resource allocation describes the decision of assigning concrete process tasks (i.e., enabled activity instances) to resources during process execution, adhering to run-time information (e.g., the resource availability) and previously defined allocation goals and constraints~\cite{cabanillas2015SpecificationAutomatedDesigntime}. In this work, we mainly focus on allocating tasks to human resources. Resource allocation can have various \emph{(optimization) goals}, such as balancing workloads, or minimizing cost or cycle time. Further, resource allocation must follow \emph{hard constraints} (e.g., the separation of concerns principle or specific permission requirements) and \emph{soft constraints} (e.g., preference for resources with a particular experience), where non-adherence may be associated with a penalty \cite{pufahl2025automatic}.\lodo{Soft constraints and optimization goals kinda overlap}

To avoid lengthy manual decision-making, resource allocation can be supported by information systems, which either fully automate the decision or provide recommendations for human decision-makers \cite{zurmuehlen2004OrganizationalManagementWorkflow, russel2004workflow}. 
These systems need to be aware of knowledge for performing suitable resource allocations.
%As allocation decisions are subject to guidelines and constraints and can have a high impact on the outcome of a process in terms of quality and (time) cost\missingsrc, such support approaches aim to optimize \cite{pufahl2025automatic} \bettersrcneeded\todo{About the sources: This overlaps with the introduction, so harmonize with the sources there}. 
We define \emph{allocation knowledge} as all this (potentially implicit) knowledge relevant for making an allocation decision. 
In literature, Ihde et al.~\cite{ihde2022framework} and Pufahl et al.~\cite{pufahl2025automatic} describe that knowledge about the resource allocation problem is needed: each activity is associated with a certain resource allocation problem with certain goals (contributing to the overall process goal) and allocation constraints (e.g., for complying to specific compliance rules or allowing a certain process quality). For handling these problems, allocation mechanisms~\cite{zurmuehlen2004OrganizationalManagementWorkflow} can be selected and are part of the allocation knowledge. This includes allocation rules as collected by Russel et al.~\cite{russel2004workflow}, such as \emph{role-based allocation}, i.e., requiring assigned resources to possess specific roles or assigning based on previous work on the case, etc.
Ouyang et al.~\cite{ouyang2010ModellingComplexResource} and Arias et al.~\cite{arias2018TaxonomyHumanResource} discuss knowledge about individual resources such as their roles, their expertise, workload, social context and the organizational setting as well as their previous performance. 
Additionally, resource behavior, such as prioritization, handovers, or batching is summarized by Rubensson et al.~\cite{rubensson2025ConceptualFrameworkResource}. 
Finally, Pufahl et al.~\cite{pufahl2025automatic} highlight that knowledge about the tasks to assign is also relevant, such as resource requirements,  priority, expected performance, and its history. 
Based on these related work, we can observe that existing research selectively consider certain allocation knowledge, but no overall knowledge categorization for resource allocation exist.
%Existing works on resource allocation already selectively propose considering further aspects\NM{Further than ...?} of allocation knowledge such as a taxonomy of human abilities for allocation decisions \cite{erasmus2018MethodEnableAbilityBased}. 
Related to our work, Havur et al. \cite{havur2016ResourceAllocationDependencies} use answer set programming, a logical reasoning approach, for resource allocation based on a knowledge graph encoding role and availability constraints. 
However, their work does not provide a framework for adapting the knowledge graph and applying explainable reasoning techniques.

When information systems or even BPMSs support the execution of business processes, process execution data is stored. Using this data in the form of event logs, \emph{process mining} offers data-driven techniques to discover the process, check its conformance, and enhance it with additional information~\cite{dumas2023AIAugmentedBusinessProcess}. 
Process mining offers a set of techniques to derive knowledge on behavioral and structural aspects of resources in business processes as reviewed in Martin and Beerepoot~\cite{martin2024UnveilingUseCases} and Rubensson et al.~\cite{rubensson2025ConceptualFrameworkResource}. 
These techniques provide opportunities to derive structural knowledge for resource allocation from process execution data.
% cp. \cref{sec:input}

%\LP{Please add: "Ouyang et al. \cite{ouyang2010ModellingComplexResource} provide conceptual data models for structuring knowledge with regards to resource allocation in business processes defining data structures with regards to resources and their classification, the resource assignment to tasks, availability, and information logging regarding resource utilization. These data models should support the design and analysis of resource-aware business processes and their implementation."}

% \subsubsection{Resource Allocation Knowledge}
%\begin{outline}
%    \1 Luise SLR \cite{pufahl2025automatic}
%    \1 Arias taxonomy \cite{arias2018TaxonomyHumanResource}
%    \1 Niels \& Iris Paper \cite{martin2024UnveilingUseCases}
%    \1 Rubensson \cite{rubensson2025ConceptualFrameworkResource}
%    \1 Ouyang paper \cite{ouyang2010ModellingComplexResource}
%\end{outline}

\subsection{Knowledge Graph Reasoning}
% \iodo{Danger zone: This background was adopted from \cite{bein2025KnowledgeGraphsKey} be careful not to get a self plagiarism. Maybe somebody else check this}
\emph{Knowledge graphs} (KG) can be defined as \quotes{graph[s] of data intended to accumulate and convey knowledge of the real world, whose nodes represent entities of interest and whose edges represent relations between these entities} \cite{hogan2022KnowledgeGraphs}.
%\emph{Knowledge graphs} (KG) can be defined as graphs of data whose nodes and edges represent entities of the real world and their relations\cite{hogan2022KnowledgeGraphs}. 
KGs additionally allow the definition of context, constraints, and inference rules in the form of ontologies \cite{hogan2022KnowledgeGraphs}. 
This context can be used to formally define the domain, e.g., in terms of types of entities and properties of relations such as source or target type limitations, symmetry, transitivity, or attribute ranges \cite{hogan2022KnowledgeGraphs}. 

This umbrella notion of a knowledge graph allows encoding a wide range of knowledge and thus a usage in diverse use cases, such as for product recommendations, using the graph to encode product and user relations \cite{hogan2022KnowledgeGraphs, wang2019ExplainableReasoningKnowledge}, or for personalized medicine, using the graph to encode biochemical interactions \cite{ruiz2021IdentificationDiseaseTreatment}.

% Formally, the data in knowledge graphs can be expressed as triples of the form $(subject, predicate, object)$, where the subject and object are entities, defined by a unique id, and the predicate is an id for a relation, usually non-unique. 
% More complex graph traits, such as node attributes, n-ary relationships, and meta-information, can be expressed this way using attribute relations, literal nodes, and relational nodes \cite{hogan2022KnowledgeGraphs}. 
% Assuming uniqueness, these triples can be expressed as first-order-logic literals in the form of $predicate(subject, object)$, which allows formal reasoning about them. 
% In addition, in this paper, we will use visualizations as exemplified by \cref{fig:graph_example_1}. For simplification, we use node labels in the form \quotes{id:typename} which imply the relation $type(id,typename)$. 

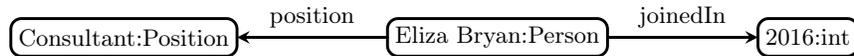
\begin{figure}[]
    \vspace{-2mm}
    \centering
    \begin{tikzpicture} 
        [
            %every node/.style = {font=\tiny},
            node distance=1mm and 20mm,
        ]
        % First graph
        \node (a) [entity] {Consultant:Position};
        \node (b) [entity, right = of a] {Eliza Bryan:Person};
        \node (c) [entity, right = of b] {2016:int};
        
        \draw (b) edge[arrow, "position" '] (a);
        \draw (b) edge[arrow, "joinedIn"] (c);
    
    \end{tikzpicture}
    \vspace{-2mm}
    \caption{Knowledge graph example. Node labels include types for simplification. }
    \label{fig:graph_example_1}
    \vspace{-5mm}
\end{figure}

Formally, knowledge graphs can be expressed as sets of triples of the form \emph{(subject, predicate, object)}, where the subject and object are unique ids representing entities, and the predicate identifies a, usually non-unique, relation between them. 
In this paper, we use graph visualizations as exemplified by \cref{fig:graph_example_1}, where the tail, label, and head of an edge represent a subject, predicate, and object, respectively. For simplification, we further use node labels in the form \quotes{id:typename} which imply the relation $(id,\texttt{type}, typename)$.

Working with knowledge graphs can be categorized into knowledge extraction, ontology engineering, and knowledge graph reasoning\cite{hogan2022KnowledgeGraphs}. 
\emph{Knowledge extraction} refers to the act of identifying the relevant entities and their relations based on data in unstructured or structured form. When integrating extracted knowledge into an existing graph, knowledge fusion needs to be performed. 
\emph{Ontology engineering} helps to structure and add meaning to the extracted knowledge.   
It encompasses methods ranging from manual elicitation to (semi-)automated learning approaches.
\emph{Knowledge graph reasoning} (KGR) aims to derive new knowledge from existing graphs, i.e., mainly new entities and especially new relations \cite{chen2022OverviewKnowledgeGraph,delong2024NeurosymbolicAIReasoning}. 
AI advances with subsymbolic methods have yielded new neuro-symbolic approaches for KGR, which integrate symbolic logic-based reasoning techniques with subsymbolic neural-net-based methods\cite{delong2024NeurosymbolicAIReasoning}. 
Consequently, many of these approaches provide explainability \cite{tiddi2022KnowledgeGraphsTools,chen2022OverviewKnowledgeGraph}. 

%\todo{Consider explaining explainability briefly}
% \todo{Put a delineation to ontologies here (also to more easily delineate from ontological bp modeling}
%\todo{Also think about talking about RDF(S), OWL, etc., this might get confusing otherwise (at least RDF?)}
%\todo{Maybe put one sentence about information retrieval to connect to related work - maybe not}
%\todo{Introduce on sentence on knowledge fusion to be referenced later?}

% \optnewpage
% \subsection{Knowledge Graphs in Business Process Management}
\subsection{Ontology-based Business Process Modeling}
\emph{Ontology-based Business Process Modeling} (OBPM), also called \emph{Semantic Business Process Modeling}, encompasses approaches to extend classical process models with instances of ontologies that represent additional background process knowledge \cite{corea2021OntologyBasedProcessModelling}. 
This semantic underpinning promises to provide missing context to model users and to allow for applying advanced reasoning techniques, e.g., for validation and compliance checking of process models \cite{corea2017DetectingComplianceBusiness,annane2019BBOBPMN20,bachhofner2022AutomatedProcessKnowledge,difrancescomarino2008ReasoningSemanticallyAnnotated}. 
Existing works show how ontology languages such as OWL can be used to express a subset of the BPMN language \cite{annane2019BBOBPMN20, bachhofner2022AutomatedProcessKnowledge} as well as to express event-log-like process execution traces \cite{annane2019BBOBPMN20}. 
While some of the works already apply ontology-based reasoning methods to instances of the modeled ontologies (cp. \cite{corea2017DetectingComplianceBusiness}), these works are limited to reasoning at process design-time and the major focus lays on modeling and integrating with existing process modeling approaches rather than reasoning \cite{corea2021OntologyBasedProcessModelling}. 
In contrast, we consider process run-time, when allocation decisions happen, and put an emphasis on automated reasoning.

%\todo{What about, e.g., proposals to integrate domain knowledge as in \cite{difrancescomarino2008ReasoningSemanticallyAnnotated}?}
%\todo{Mention the requirements for ontological PMod from \cite{corea2021OntologyBasedProcessModelling}?}

\optnewpage

\section{A Knowledge-Graph-Based Resource Allocation Framework} \label{sec:main}
\begin{figure}[htbp]
  \vspace{-3mm}
  \centering
  \includesvg[width=.9\columnwidth, pretex=\relscale{0.85}]{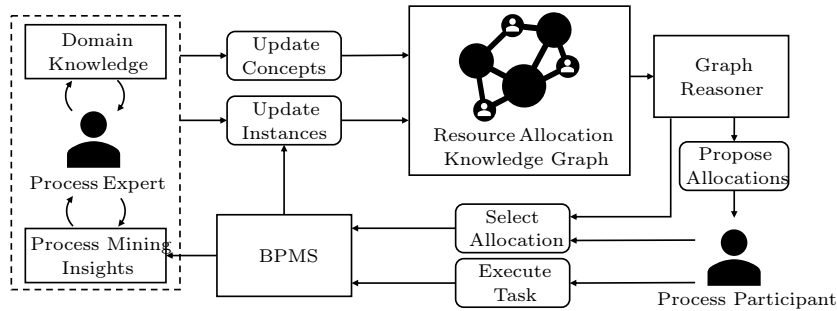}
  % \includesvg[width=\columnwidth]{figures/overview.svg}
  \caption{Conceptual overview of KRAFT, a Knowledge-graph-based Resource Allocation Framework for Tasks}
  \label{fig:overview}
  \vspace{-3mm}
\end{figure}

%
\begin{comment}
\begin{outline}
\todo{Recheck later to align with motivation earlier in the paper (or extend if not sufficiently done before)}
    \1 As shown in \cite{cabanillas2016ProcessResourceAwareInformation}, current BPMS, i.e., resource allocation support systems, are limited and rigid w.r.t. the assignment conditions that they allow to specify. 
    \1 In order to allow to be more flexible in the assignment mechanisms specifiable, the knowledge to be considered, and the adaptability of said mechanisms and knowledge,
    \1 we propose borrowing from ontological business process modeling and knowledge graph reasoning.
    \1 
\end{outline}    
\end{comment}
%
This section presents \emph{KRAFT, a Knowledge-graph-based Resource Allocation Framework for Tasks}, that integrates organizational process mining, ontology-based allocation knowledge modeling, and knowledge graph reasoning, to deliver suitable and explainable resource allocations in processes. \Cref{fig:overview} provides an overview of KRAFT.

% \subsection{\sout{Framework and Architecture} Overview}

% \todo{Fig.: Should there be a connection from BPMS to graph reasoner to know when an allocation decision is necessary}%
% \todo{Fig.: Minor inconsistency: Domain Knowledge, DTD, and AKG are artifacts while BPMS and GR are rather systems. Reevaluate.}%
%

% \NM{New version of the introductory paragraph of KRAFT - please review. The old version is still present, but commented out. LB: Noice}
At the heart of KRAFT, there is the \emph{resource allocation knowledge graph} that encodes the relevant entities and their relations, as well as, at an abstracted level, the organization's allocation concepts and mechanisms. This graph is populated in two ways. First, the \emph{process expert}, i.e. the person responsible for defining the organization's allocation mechanisms, can update the graph both conceptually, e.g., by introducing new allocation mechanisms, and at an instance level, e.g., by adding new employees or changing roles. To this end, the process expert can leverage a combination of process mining insights and domain knowledge. Second, the \textit{BPMS} (potentially also KG-based \cite{bein2025KnowledgeGraphsKey}) can update the graph with execution information such as running tasks. 
Using the resource allocation knowledge graph, \emph{reasoning approaches} are applied to infer potential allocations for currently open tasks, together with an assessment and reasoning for each potential allocation. The \emph{best proposals} can either be (i)  automatically allocated or (ii) reviewed by the process participants to make an allocation decision. By executing the allocated tasks, the process state gets updated, which the BPMS forwards to the graph and reasoner by updating the respective information.

\optnewpage
\subsection{Resource Allocation Knowledge Graph} \label{sec:schema}
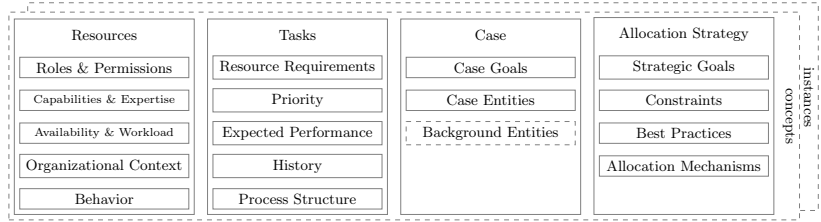
\begin{figure}[h]
    \vspace{-3mm}
    \centering
    \resizebox{.9\columnwidth}{!}{
    \tikzstyle{arrow} = [thick,->,>=stealth, line width=1.5pt]
    \begin{tikzpicture}[
      node distance=7mm and 1cm,
      title/.style={minimum width=3.6cm},
      activity/.style={rectangle, draw=black!50, anchor=west, minimum width=3.6cm}
    ]
    \node (phase1) [title] {Resources};
        \node (1_1) [below=of phase1.west, activity, xshift=0mm] { Roles \& Permissions} ;
        \node (1_2) [below=of 1_1.west, activity, font=\relscale{.87}] { Capabilities \& Expertise} ;
        \node (1_3) [below=of 1_2.west, activity, font=\relscale{.87}] { Availability \& Workload} ;
        \node (1_4) [below=of 1_3.west, activity] { Organizational Context} ;
        \node (1_5) [below=of 1_4.west, activity]  { Behavior};
        %\node (1_6) [below=of 1_5.west, activity, opacity=0]  { Placeholder };
        %\node (1_7) [below=of 1_6.west, activity, opacity=0]  { Placeholder };
    \node (p1) [draw=black!50, fit={(phase1) (1_1) (1_2) (1_3) (1_4) (1_5) %(1_6) (1_7)
    }] {};
    
    \node (phase2) [right = 0.5cm of phase1, title] {Tasks};
        \node (2_1) [below=of phase2.west, activity, xshift=0mm] { Resource Requirements} ;
        \node (2_2) [below=of 2_1.west, activity] { Priority} ;
        \node (2_3) [below=of 2_2.west, activity] { Expected Performance} ;
        \node (2_4) [below=of 2_3.west, activity] { History} ;
        \node (2_5) [below=of 2_4.west, activity] { Process Structure} ;
        %\node (2_6) [below=of 2_5.west, activity, opacity=0] { Placeholder };
        %\node (2_7) [below=of 2_6.west, activity, opacity=0] { Placeholder };
    \node (p2) [draw=black!50, fit={(phase2) (2_1) (2_2) (2_3) (2_4) (2_5) %(2_6) (2_7)
    }] {};

    \node (phase3) [right = 0.5cm of phase2, title] {Case};
        \node (3_1) [below=of phase3.west, activity, xshift=0mm] { Case Goals } ;
        \node (3_2) [below=of 3_1.west, activity] { Case Entities } ;
        \node (3_3) [below=of 3_2.west, activity, dashed] { Background Entities };
        \node (3_4) [below=of 3_3.west, activity, opacity=0] {  };
        \node (3_5) [below=of 3_4.west, activity, opacity=0] { Placeholder };
        %\node (3_6) [below=of 3_5.west, activity, opacity=0] { Placeholder };
        %\node (3_7) [below=of 3_6.west, activity, opacity=0] { Placeholder };
    \node (p3) [draw=black!50, fit={(phase3) (3_1) (3_2) (3_3) (3_4) (3_5) %(3_6) (3_7)
    }] {};
    
    \node (phase4) [right = 0.5cm of phase3, title] {Allocation Strategy};
        \node (4_1) [below=of phase4.west, activity, xshift=0mm] { Strategic Goals } ;
        \node (4_2) [below=of 4_1.west, activity] { Constraints} ;
        \node (4_3) [below=of 4_2.west, activity] { Best Practices } ;
        \node (4_4) [below=of 4_3.west, activity] { Allocation Mechanisms };
         \node (4_5) [below=of 4_4.west, activity, opacity=0] { Placeholder };
        % \node (4_6) [below=of 4_5.west, activity, opacity=0] { Placeholder };
        % \node (4_7) [below=of 4_6.west, activity, opacity=0] { Placeholder };
    \node (p4) [draw=black!50, fit={(phase4) (4_1) (4_2) (4_4) (4_4) (4_5) %(4_6) (4_7)
    }] {};

    \node (s_label) [label={[rotate=-90]center:concepts}, right=2mm of $(p4.east)$] {};
    \node (d_label) [label={[rotate=-90]center:instances}, right= 2mm of s_label, yshift=4mm] {};
    \begin{scope}[on background layer]
        \node (dynamic) [draw=black!50, dashed, fit={(p1) (p4) (s_label)}, yshift=2mm, xshift=4mm] {};
        \node (static) [draw=black!50, dashed, fit={(p1) (p4) (s_label)}, fill=white] {};
    \end{scope}
    
    \end{tikzpicture}
    }
    \vspace{-2mm}
    \caption{Overview of resource allocation knowledge, extended from \cite{ouyang2010ModellingComplexResource,arias2018TaxonomyHumanResource,pufahl2025automatic}}
    \vspace{-3mm}
    \label{fig:allknowledge}
\end{figure}

The resource allocation knowledge graph is aimed at making all necessary allocation knowledge available for reasoning whilst supporting easy adaptability. 
In the following, we provide our classification of the \emph{allocation knowledge} with a focus on human resources to be encoded in the graph,  
leveraging and extending the literature presented in \cref{sec:resall}. \Cref{fig:allknowledge} visualizes this classification, distinguishing resource, task, case and allocation strategy knowledge.

\emph{Resource knowledge} covers knowledge about the resources as main entities for allocation. 
This includes knowledge 
regarding the individual roles and permissions (determining which tasks a resource is responsible for and permitted to do), 
the individual capabilities and expertise (determining the resource's aptitude for tasks and consequently its performance), 
the resource's availability (e.g., working hours) and workload (e.g., general amount of tasks being handled), 
the organizational context (i.e., where resources are based, which teams they belong to, etc.), 
and -- often implicit -- behavior (e.g., prioritization).

\emph{Task knowledge} pertains to knowledge about single tasks, i.e., instances of activities. 
This groups resource requirements related to a particular task, e.g., defined based on the resource's knowledge and experience, as well as other task-related factors such as the tasks's priority and expected performance. 
Further, it includes the aggregated history of previously executed and currently running tasks. 
Lastly, we count the structure of the process, which encodes the relations between the activities, in this category. 

\emph{Case knowledge} refers to the knowledge about cases, i.e., instances of processes, including individual case goals and entities. 
Notably, we introduce the notion of \emph{background entities} and concepts, which has only sparsely been touched in literature. Under this subcategory, all the relevant domain-specific entities and concepts related to a case are collected. For example, in a medical process, this would relate to, e.g., medical concepts and their interrelations, connected to the patient case. In a loan application process, this would relate to, e.g., risk classes and loan goal types attached to a loan application case. 

Lastly, we introduce the notion of \emph{allocation strategy knowledge}. 
Strategy is influenced by organizational goals exceeding individual case goals, e.g., \quotes{the workload should be distributed equally}. 
Further, \emph{constraints}, stemming from regulations or organizational \emph{best practices}, are collected here. 
Finally, \emph{allocation mechanisms} refers to knowledge about the methods applied for making allocation decisions, such as allocation rules (e.g., separation of concerns).% and patterns like .

% \2 describe what kind of instances can be present in the graph and what kind of inferences can be made
% \2 This is equivalent to the T-box and R-box in description logic vocabulary \todo{Later crosscheck with background}
%For example, in case of a loan application process, concept knowledge would on the one hand detail, e.g., that there are activities and resources, as well as loan goals and risk classes. 
%On the other hand, it would detail how these relate and what they mean, e.g., that tasks are instances of activities, or that certain risk classes imply the need for more senior clerks. 
%

It is further useful to divide allocation knowledge along another axis into knowledge about \emph{concepts} and \emph{instances}. 
%\NM{If the introduction of these terms does not move upward to the background section, we should also check the introductory paragraph of KRAFt as these terms are also used there. LB: Changed the wording slightly}
Under concept knowledge, analogously to KG ontologies, we group knowledge about what kind of entities can exist, how they relate, what they mean, and how allocation reasoning can be performed on them. 
% \NM{I like the example, but be aware that we use 'instance' quite often with different interpretations. LB: Yes. It's an overloaded term from the beginning on, but then task and cases are instances of activities and processes while at the same time all of them are instances of their respective classes} 
For example, in a loan application process, concept knowledge includes that there are activities, resources, loan goals, and risk classes, as well as that tasks are instances of activities or that certain risk classes imply the need for more senior clerks. 
On the other hand, we consider the knowledge that there is a specific activity \quotes{Check for Fraud}, a task \quotes{t576} that is an instance of this activity, and an encompassing loan application case \quotes{c56} referencing the loan goal of \quotes{Financing a Car}, as knowledge about instances of these concepts. %, but also that there is a separation of concerns in place for activities \quotes{Validate Application} and \quotes{Assess Potential Fraud}. 
To implement the resource allocation knowledge graph, common knowledge graph technologies can be utilized, such as standardized ontology languages like OWL. 
Further, following guidelines from ontology-based process modeling \cite{corea2021OntologyBasedProcessModelling}, a respective system implementing the resource allocation knowledge graph should provide an initial set of standard concepts to organizations, which captures the most common allocation knowledge as described above. 
Organizations can build upon what they need from this knowledge and extend it with their individual concepts. 
Existing ontologies of organizational knowledge, such as the \emph{W3C Organization Ontology}\footnote{\url{https://www.w3.org/TR/vocab-org/}} or the \emph{BPMN-Based Ontology}\footnote{\url{https://www.irit.fr/recherches/MELODI/ontologies/BBO}, see also \cite{annane2019BBOBPMN20}} can be taken as foundation to build this standard concept graph. 
However, such existing ontologies conflict with each other, e.g., in how to represent roles, and cover the dimensions described above only insufficiently, either focusing on resources or task and process knowledge and mostly ignoring allocation strategy knowledge.

\optnewpage
\subsection{Knowledge Extraction \& Graph Population}
%\todo[inline]{Add an introductory sentence here?}
\subsubsection{Knowledge Extraction} 
Relevant information must be extracted as a first step in populating or updating the knowledge in the graph. We identify two main sources that organizations can leverage: (1) process mining insights and (2) domain knowledge. 
\emph{Process mining techniques} can extract resource allocation knowledge from historical process data captured in event logs. %when executing processes using information systems and BPMSs. 
These techniques are particularly effective in identifying knowledge related to resources, e.g., roles, expertise, and behavior\cite{martin2024UnveilingUseCases,rubensson2025ConceptualFrameworkResource}. Additionally, they can extract task-specific knowledge, such as process structures~\cite{dumas2018FundamentalsBusinessProcess}, resource requirements, and case-specific knowledge, including case entity discovery~\cite{rubensson2025ConceptualFrameworkResource}. Furthermore, historic allocation mechanisms~\cite{pufahl2025automatic} can be inferred. 
We assume that a process expert utilizes a process mining tool supporting such techniques, e.g., PM4Py~\cite{berti2023pm4py}. Based on the insights obtained from process mining analysis, the expert selects and integrates relevant knowledge into the graph, as described below.

While process mining excels at extracting resource- and task-related knowledge, \emph{domain knowledge} is essential for aspects that cannot be directly observed in execution data, such as allocation strategies, strategic goals, constraints, and best practices. 
Domain knowledge can take different forms: it may exist as implicit expert knowledge, which can be elicited through a conversational interface, or as explicit data. This data may be unstructured (e.g., regulatory documents) or structured, originating from existing systems managing allocation knowledge. 

Different extraction techniques are applicable based on the knowledge source and format. For textual sources, including human input, Natural-Language-Processing-based information extraction is effective for knowledge graph construction \cite{hogan2022KnowledgeGraphs}. Recent advances in large language models 
further enhance these capabilities, making them a promising tool for structuring domain knowledge \cite{zhu2024LLMsKnowledgeGraph}.

\subsubsection{Graph Population} \label{sec:input}
\begin{figure}[h]
  \vspace{-7mm}
  \centering
  \resizebox{.85\columnwidth}{!}{\includesvg[width=1.15\columnwidth]{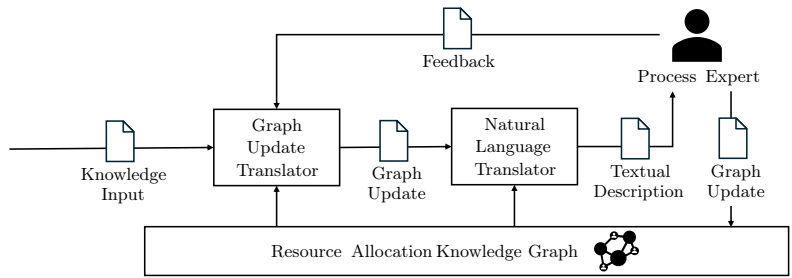}} % The 1.15 fixes the text
  \caption{Conceptual overview of knowledge input including feedback mechanism}
  \label{fig:input}
  \vspace{-5mm}
\end{figure}

\begin{comment}
In a next step, extracted knowledge needs to be integrated into the graph. 
This includes, on one hand, updating concepts, i.e., what kind of entities can be present in the graph, what they mean, and how reasoning can be performed on them, which is done by the process expert. 
On the other hand, the concrete entities, i.e., instances of these concepts, are updated by the process expert and the BPMS. 
\end{comment}
In a next step, the extracted knowledge needs to be integrated into the graph. 
As described above, this includes updating concepts by the process expert and updating instances by the process expert and the BPMS. \lodo{Reference to concepts}
For updating run-time instances by the BPMS, an appropriate bridge between the two systems must be implemented. 
For new updates from the process expert, we envision the workflow shown in \cref{fig:input}. 

As most knowledge extraction methods do not directly produce formal graph updates, extracted knowledge must first be translated into structured changes that align with the knowledge graph, i.e., \emph{graph updates}. 
Subsequently, especially for unstructured sources such as textual inputs or expert feedback, human validation is necessary, to avoid introducing errors into the graph.

To ensure accuracy, we propose a \emph{conversational human-in-the-loop approach}. Graph updates are first converted into a human-readable format for review, leveraging ontological meta-information stored in the graph, such as class descriptions. If the process expert provides corrections, the system refines its update and presents a revised version for validation. Once confirmed, the update is applied to the graph, and the system learns from the interaction. 
For example, consider a process expert adding a new employee by typing: \quotes{Eliza Bryan has joined us as a senior consultant}. 
The system translates this into a graph update and then back into text: \quotes{I will add a new person called \squotes{Eliza Bryan} with the role \squotes{Senior Consultant}}. 
The expert corrects it by stating: \quotes{No, she will take the role of \squotes{Consultant} but with the seniority \squotes{Senior}\,}. 
The system refines the update accordingly and applies it upon confirmation, learning that seniority can be an attribute of existing roles rather than a separate role.

Automating this process requires NLP techniques, for summarizing and refining graph updates. In particular, large language models seem highly suitable. However, deriving valid updates while incorporating human feedback is complex and beyond the scope of this paper. Instead, we focus on the overall framework for knowledge-graph-based resource allocation, while future work may explore knowledge fusion techniques to enhance this integration.

\optnewpage
\subsection{Allocation Reasoning}
Goal of the allocation reasoning is to consider the knowledge encoded in the graph, assess potential allocations, and provide explanations for the assessments. 
Following the different degrees of automation, the component must both allow automatically selecting an allocation and letting a human select from a set of best options with explanations. 
In case human decision-making strongly deviates from system proposals, self-improvement using a mechanism similar to the one described in \cref{sec:input} appears sensible. 
%\todo[]{Put an example?}

% \paragraph{Technologies}
The internal workings of the reasoning strongly depend on the technologies utilized. 
Knowledge graph reasoning (KGR) and adjacent fields provide a wide toolbox of approaches to achieve explainable recommendations (cp. \cite{delong2024NeurosymbolicAIReasoning, hogan2022KnowledgeGraphs}). 
Specifically, we can interpret allocation as the problem of \emph{predicting links} between resource and task entities, one of the most common use cases of KGR. 
Notably, KGR approaches often combine multiple technologies, borrowing from both symbolic and subsymbolic methods to varying degrees, placing them as neuro-symbolic \cite{delong2024NeurosymbolicAIReasoning}. 
When selecting appropriate methods, a tradeoff has to be made. 
Subsymbolic approaches tend to be more efficient, generalizable, and more suitable for continuous target variables. % Note: Yes, the rather is necessary, as it's not about completely either or 
Conversely, symbolic approaches are more comprehensible, more stable to change, and allow for better explainability \cite{tiddi2022KnowledgeGraphsTools, hogan2022KnowledgeGraphs, wang2019ExplainableReasoningKnowledge,chen2022OverviewKnowledgeGraph}. 
For instance, one family of neuro-symbolic approaches learns and considers weighted path or graph pattern rules for link prediction, providing the deciding rules as fairly interpretable output along with the edge \cite{delong2024NeurosymbolicAIReasoning, hogan2022KnowledgeGraphs, chen2022OverviewKnowledgeGraph}.

Specifically for resource allocation, on the symbolic side, description logic ontologies in combination with answer set programming have already been utilized for compliance checking\cite{corea2017DetectingComplianceBusiness} and scheduling under role and availability constraints \cite{havur2016ResourceAllocationDependencies}. 
On the subsymbolic side, neural nets have already been used to predict execution times based on process execution traces\cite{verenich2019SurveyCrossbenchmarkComparison}, which should be also applicable to graph-based inputs. 
As an exemplary combination, these runtime estimations could be utilized in rule-based reasoning.
%\todo{Extend a bit more on what technologies might be valuable here (?)}

%\todo{Could also add (commented out) sth. about potential technologies or challenges for this matching}

\begin{comment}
    
{\color{blue!60!red}
%    \1 Many techniques for graph reasoning are available more or less out of the box

\paragraph{Potential approaches}
\begin{outline}
    \1 Symbolic: (can usually be extended by subsymbolic mining approaches)
        \2 Pattern-based or path-based rules
        \2 (Graph analytics)
    \1 Subsymbolic
        \2 Graph embeddings
        \2 
\end{outline}

\paragraph{Challenges}
\begin{outline}
    \1 Difficulty: Classical knowledge graph problem of graph completion doesn't 100\% match a continuous variable prediction problem
        \2 Difficulty: Long-term planning etc. 
        \2 <=> This rather applies to optimization, which we don't really do anymore
\end{outline}
}

\end{comment}

\optnewpage
\section{Proof of Concept} \label{sec:poc}
This section outlines our prototypical implementation of KRAFT and demonstrates the general feasibility and usefulness of knowledge-graph-based resource allocation support. 
%\todo{Mention that this focuses on reasoning over ontology engineering and knowledge extraction? }

\subsection{Illustrative Setting} \label{sec:uc}
\begin{figure}[h]
  \vspace{-5mm}
  \centering
  \includesvg[width=0.75\columnwidth, inkscapelatex=false]{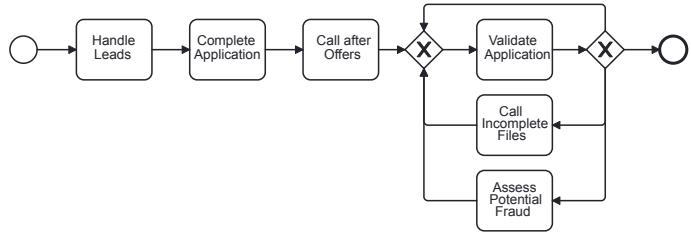} 
  \vspace{-3mm}
  \caption{Main flows for the example loan application process}
  \label{fig:processModel}
  \vspace{-5mm}
\end{figure}

As a synthetic demonstration case for our prototype, we assume a loan application process in a financial institution. 
This process is based on an adapted form of the Business Process Intelligence Challenge 2017 dataset\cite{vandongen2017bpic}, as provided by the Business Process Optimization Challenge 2023\footnote{See \url{https://sites.google.com/view/bpo2023/competition}} (BPOC). 
The dataset has been cleaned and aggregated w.r.t. the activity lifecycle. 
\Cref{fig:processModel} represents the resulting main process flows as BPMN diagram.
% \todo{Consider removing the empty loop again, as the previous activities also could loop}

Cases of this process have two categorical attributes, \emph{Application Type} and \emph{Loan Goal}, as well as one numerical attribute, the \emph{Requested Amount}. 
In addition, we define further allocation knowledge and associated mechanisms. This includes \emph{Risk Class}es attached to loan goals and \emph{Seniority} attached to resources, where higher risk classes require higher seniority when working on a case, as well as \emph{Separation of Concerns} (ensuring different resources handle a specific set of activities). Further, we added individual \emph{Expertise} for specific case attributes. 
To substitute a real-world BPMS, a business process simulation frame adapted from the BPOC is used%s Business Process Optimization Framework\footnote{See \url{https://github.com/bpogroup/bpo-project/}}
, which 
logs relevant events and 
triggers necessary allocation decisions.
% (cp. \texttt{resource} folder in the BPOC repository for event log preprocessing)
The simulation frame requires one-to-one allocation based on provided direct resource-to-activity permissions.

\optnewpage

\subsection{Prototype Design}
\vspace{-0.5mm}
To demonstrate KRAFT, we have implemented a simple application that showcases the framework's concepts on an elementary level. As interface to the prototype, we provide a Jupyter notebook, which is prefilled with a demonstration\footnote{The prototype code and the demonstration notebook are available at
\texttt{\href{https://anonymous.4open.science/r/knowledge-graph-resource-allocation-FA43}{anonymous.4open.science/r/knowledge-graph-resource-allocation-FA43}}.
}.

The prototype's resource allocation graph is based on common knowledge graph technologies: %, as standardized by the W3C: 
all nodes and edges are encoded in the W3C Resource Description Framework (RDF) format. 
Concept knowledge\lodo{Reference to concepts} is stored using the W3C standards Web Ontology Language 2.0 (OWL) and SHACL, a W3C standard language for expressing constraints on RDF graphs. %, which internally, in turn uses the graph query language SPARQL to express the patterns. 

Knowledge extraction has been performed on the BPIC 2017 event log, applying Pika et al.'s~\cite{pika2017MiningResourceProfiles} Resource Behavior Indicators to determine resource seniority and expertise. 
To input knowledge, RDF triples can be imported from respective files or directly created with code, and then added to the respective subgraphs.

Reasoning in the prototype is based on the above-mentioned SHACL, which allows to define rules in the form of graph patterns, as well as dynamic messages and scores for pattern matches. 
During \quotes{run-time}, i.e., when running the simulation, it checks these rules and determines an aggregated score and assessment for potential assignments, which can be considered in two modes. 
In \emph{automatic} mode, the resource with the highest score is automatically selected. 
In \emph{human-in-the-loop} mode, users are prompted to select the appropriate resource. 
The prototype supports pausing the simulation to switch modes or input knowledge at run-time. 
The detailed workings of the prototype are described in the following section using demonstrative examples.

%\LP{This is a bit tough to follow. I think that it would be helpful to structure into the points of the framework: 1) RA knowledge graph, 2) graph population, and 3) allocation reasoning--> how has each of these functionalities be implemented, how is the simulator connected to this implemented functionalities; the reasoning is missing! \\
% LB: In order to avoid too much duplication, I propose and realized the following: Make 4.2 focus solely on technical details; put all the relation to the framework into the demonstration, so there is on straight storyline walking through the three framework points}

%Each rule produces a potentially dynamic score and a message if its pattern is fulfilled. 
%The instance-level knowledge is stored as simple RDF graph. \todo{Should you really be throwing around with acronyms here? \\ Make sure things are introduced in background as necessary}
% Note, too, that, e.g., role-based allocation, is not used in the example use case. 
% {\color{blue}
%As of now, the prototype provides only very rudimentary support for schema extension in the form of an %endpoint to add new OWL and SHACL code. 
%}

\optnewpage

\vspace{-0.5mm}
\subsection{Demonstration}
\vspace{-0.5mm}
% \iodo{Definitely mention the "process mining" performed}
% \todo{\sout{Put example knowledge adaption here (otherwise adaptability requirement cannot be demonstrated)} \\ Note: The system already integrates new resources as soon as they appear in the simulation frame - maybe that's also interesting?}%
%
%
We can use this prototype to integrate allocation standard concepts as well as the allocation knowledge described in the example case of \cref{sec:uc}. 
%\LP{Do we have in the implementation somewhere a bigger picture of the graph available that we could reference here?}
Following the guidance from \cref{sec:schema}, tasks, activities, cases, resources, and roles, as well as direct and role-based allocation are part of the base concept graph\lodo{Reference to concepts?} of the prototype. 
In contrast, the domain-specific concepts of the example case, such as resource seniority, application types, and risk-class-based allocation, are use-case-specific extensions of this schema. 
For an interactive visualization of the resulting graph structure, we refer to the repository. 

Considering the example case, at some point the company added this non-standard knowledge in the form of domain-specific entities and allocation rules. In the prototype, this would be expressed in the form of OWL and SHACL structures to be added. 
For instance, adding loan goals, attached risk types, and the respective allocation rule encompasses adding classes \texttt{LoanGoal} and \texttt{RiskType} and their respective relations \texttt{hasLoanGoal} and \texttt{hasRiskType} to the ontology. 
Further, a SHACL rule defining the graph pattern for the allocation rule has to be added. \Cref{fig:rcpattern} visualizes the pattern as a graph\footnote{For ease of reading, we have refrained from putting SHACL code into the paper. We refer interested readers to the code repository.}.
%
%
%\begin{figure}

%\begin{wrapfigure}[12]{r}{0.5\columnwidth}
\betterwrapfig{0.48\textwidth}{11}{
    \centering
    \resizebox{\textwidth}{!}{
    \begin{tikzpicture} 
        [
            node distance=7mm and 10mm,
            edge quotes/.style = {sloped},
            every node/.style = {font=\tiny}
        ]
        \node (S1) [entity] {s1:Seniority};
        \node (R) [entity, below = of S1] {r:Resource};
        \node (T) [entity, below = of R] {t:Task};
        \node (PI) [entity, right = of T, xshift=-3mm] {c:Case};
        \node (LG) [entity, right = of PI, xshift=3mm] {lg:LoanGoal};
        
        \node (RC) [entity, above = of LG] {rc:RiskClass};
        \node (S2) [entity, above = of RC] {s2:Seniority};
        \coordinate[below = of S2] (X);
        
        \draw (T) edge["performedBy"', arrow, near end] (R);
        \draw (T) edge[arrow] node[yshift=3mm] {partOf} (PI);
        \draw (PI) edge[arrow] node[yshift=3mm] {hasLoanGoal} (LG);
        \draw (LG) edge["hasRiskClass", arrow, near end] (RC);

        \draw (RC) edge["minSeniority", arrow] (S2);
        \draw (R) edge["seniority"', arrow] (S1);
        \draw (S1) edge["greaterEq", arrow] (S2);
    \end{tikzpicture}
    }
    \vspace{-5mm}
    
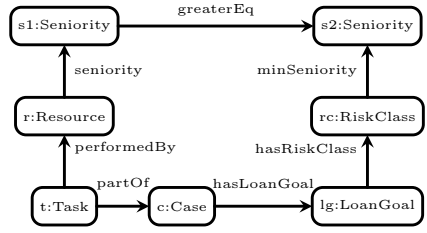
\captionof{figure}{Risk class based seniority requirement expressed as graph pattern}
    \label{fig:rcpattern}
} % Don't put anything here
The pattern, forming a simple circle, can be interpreted as follows: \emph{There is a resource} \texttt{r} \emph{performing a task} \texttt{t} \emph{for a case} \texttt{c} \emph{with a specific loan goal} \texttt{lg} \emph{and attached risk class} \texttt{rc}. \emph{The seniority} \texttt{s1} \emph{of the resource must be greater or equal than the minimum seniority} \texttt{s2} \emph{of the risk class}.
In addition to the pattern, the rule formulates a parameterizable message for matches, such as \quotes{Seniority \squotes{\texttt{s1}} is sufficient for risk class \squotes{\texttt{rc}} of loan goal \squotes{\texttt{lg}} }. 
Note that, analogous to this positive pattern that matches adherences to the rule, we also define a negative pattern matching violations, which is then used to penalize violating allocations.

\begin{figure}[h]
    \vspace{-3mm}
%\begin{wrapfigure}[10]{R}{0.5\textwidth}
%    \resizebox{.5\columnwidth}{!}{
        \centering
        \fcolorbox{white}{white}{\includegraphics[width=.7\columnwidth,trim={3.5cm 3.5cm 3.5cm 3.5cm},clip]{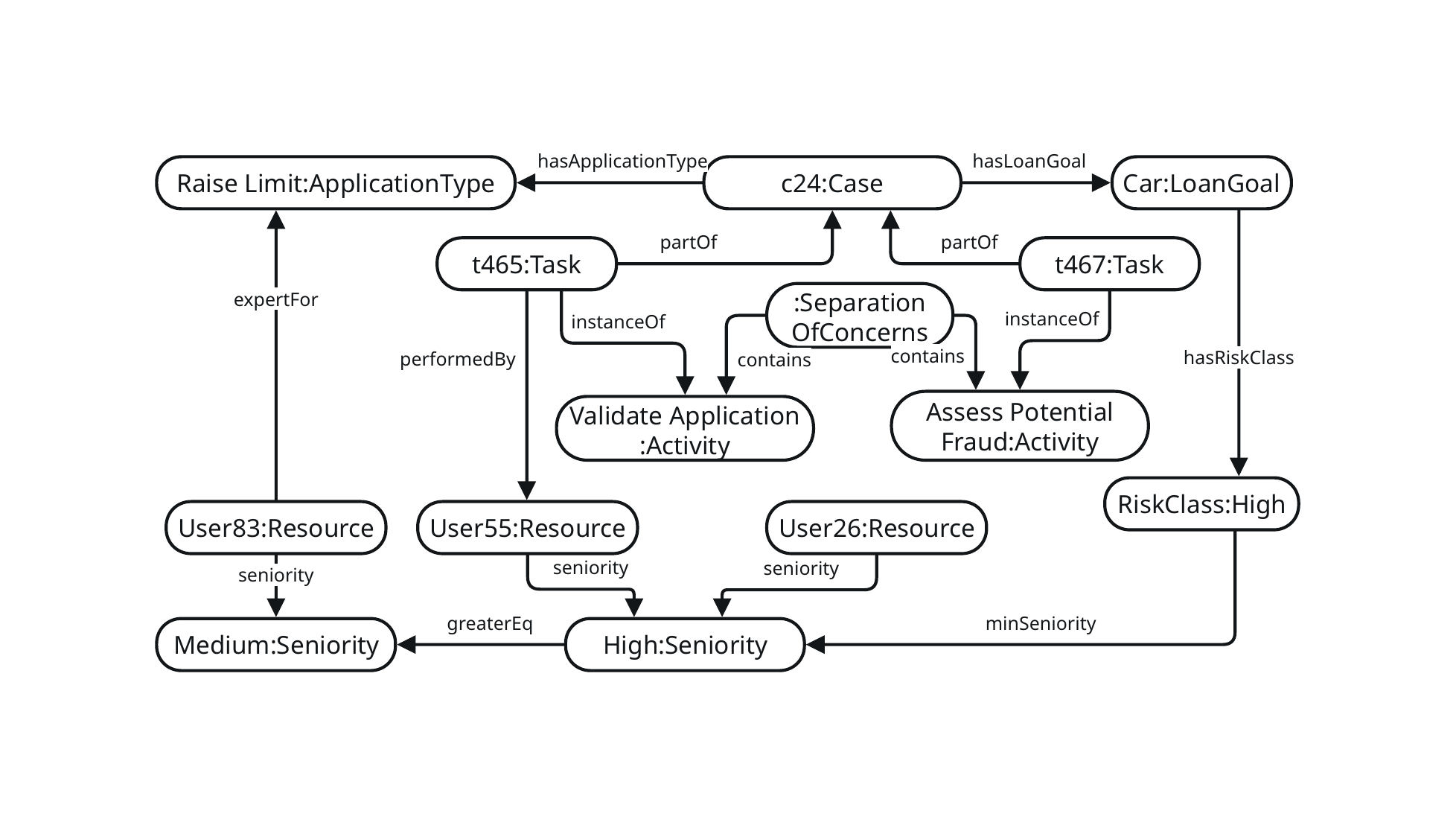}} 
%    }   
    \caption{Allocation knowledge graph excerpt with key nodes and edges for example allocation decision}
    \label{fig:graphexample}
    \vspace{-5mm}
%\end{wrapfigure}
\end{figure}

Now consider the case of a concrete application for raising the credit limit for a car: handling leads, completing the application, calling after offers, and validating the application have already been performed. Now, potential fraud should be assessed and the proper resource needs to be selected. Resources \texttt{User\_26}, \texttt{User\_55}, and \texttt{User\_83} are available and all possess roles to be allowed to perform the activity. \texttt{User\_55} has been responsible for the case so far. 
\Cref{fig:graphexample} shows an excerpt of the prototype's resource allocation knowledge graph representing relevant entities and relations for this example. 

% \todo{\Cref{fig:graphexample} should have: Previous tasks + resources + activities; case + case attributes + risk class; separation of concerns rule}

The system now tries out different assignments and considers its known rules. 
% \squotes{User\_121} does not have a role that can perform \squotes{Assess potential fraud}. 
\texttt{User\_55} has 
%the necessary permission to perform \squotes{Assess potential fraud} but has 
performed \texttt{Validate application} in the same case, which is connected to \texttt{Assess Potential Fraud} via a separation of concerns node, and thus is not eligible to perform the task. 
Assigning \texttt{User\_83} would not break the separation on concerns rule. The person is additionally an expert on the application type \texttt{Raise Limit}. However, their seniority is lower than the high seniority required for high-risk car loans. 
In contrast \texttt{User\_26} does not have special expertise on limit raises, but their high seniority matches the requested one. 
Assuming that the defined scores weigh the seniority higher than the expertise, in automatic mode, the system selects \texttt{User\_26} and logs the reasoning as shown in \cref{lst:explainability}. 
In human-in-the-loop mode, a human could overrule this weighting. 

\begin{code}
\begin{minted}[fontsize=\scriptsize,frame=single]{text}
case-1 task-7: W_Assess potential fraud 
Resources Available: {'User_26', 'User_55', 'User_83'} 
Assigning: User_26 to task-7 considering the following:
    Assignment conforms separation of concerns with activity 'W_Validate application'
    Seniority 'High' is sufficient for risk class 'High' of loan goal 'Car'
\end{minted}
\vspace{-2mm}
\captionof{listing}{Example output for on allocation decision}\label{lst:explainability} 
\vspace{-3mm}
\end{code}

\optnewpage

\section{Discussion \& Future Work} \label{sec:disc}
Based on the presented framework details and the prototypical demonstration, this section discusses KRAFT w.r.t. the objectives defined in \cref{sec:intro} and elaborates on avenues for future research. 

\subsection{Discussion}
%\todo[inline]{Glue text}
\paragraph{\cref{obj:hol} Holistic Integration.}
The prototype shows how common base allocation knowledge and domain-specific knowledge can be integrated and jointly considered for reasoning. 
The demonstration integrates various kinds of allocation knowledge (cp. \cref{fig:allknowledge}), such as resources, tasks, and cases as classes, direct allocation in the form of \texttt{canBeExecutedBy} edges between activities and resources, expertise as \texttt{expertFor} edges from resources, task history through attributes such as \texttt{startedAt} and \texttt{directlyFollowedBy} edges, and domain-specific case attributes and entities in the form of the application type, loan goal, and risk type classes. 
Further, allocation mechanisms in the form of rules, such as separation of concerns and risk-attached seniority, have been demonstrated. 
%Further, rules knowledge\NM{Has this term been used/introduced before?} has been demonstrated in the form of permissions and separation of concerns, and best practices in the form of expertise and risk-attached seniority. 
% Lacking: Workload and location; complex resource requirements, priority, expected performance, process structure, concrete goal?, goals

\paragraph{\cref{obj:adap} Easy Adaptability.}
The demonstration showed how the concept of risk-class-based allocation can be introduced with simple additions to the allocation knowledge graph. 
We further want to highlight that, for instance, changing risk classes for a specific loan goal can be easily implemented by changing one edge. In contrast, in a system that does not explicitly consider the \quotes{hidden} dependencies of allocation decisions, one would explicitly have to know to adapt the allocation mechanisms when risk classes are adjusted. 
We argue that other allocation knowledge aspects can be as easily integrated and adapted in a similar fashion as demonstrated. 
However, for extracting some of these aspects, more complex techniques might be necessary, such as predictive process monitoring techniques for expected performance. 
Moreover, a more user-friendly interface is necessary to demonstrate the ease of adaptation. 
% \todo{One could say it's easy w.r.t the knowledge you can add, but not w.r.t. how you add the knowledge?}
% \todo{Right now, the system has some flexibility in the form of lazily loading anything that comes in. Should we frame that as contribution?}

%\begin{outline}
%        \2 Nice feedback mechanism not yet implemented in the prototype, only proof of concept
%    \1 And: system allows full freedom in adding new concepts
%        \2 All of the concepts described above can be added, and all described in \cref{fig:allknowledge} could be
%\end{outline}

\paragraph{\cref{obj:expl} Explainability.}
The prototype shows a simple, rule-based implementation of explaining its allocation decisions. 
For automated decisions, reasonings are logged which allows users to understand why the system has made a decision and to make adjustments to the allocation knowledge in the graph. 
Further, in manual mode, the system provides reasoning that humans can interpret for semi-automated decision-making. Consequently, users do not have to blindly accept system decisions or manually reproduce all the reasoning, while being able to understand where wrong conclusions were made and override them.
% \LP{What just came into my mind is that explainability is also often a requirement in regulations. AI and automation techniques cannot be used in certain domains because it need to be reasoned and documented why a certain resource was selected and with our approach this is possible.}

\begin{comment}
The graph all supports:
\begin{outline}
    \1 Tasks, activities, cases
    \1 Application types
    \1 loan goals
        \2 Attached risk class \todo{Still to be impl. in simulator} 
    \1 Resources
        \2 Seniority  \todo{Still to be impl. in simulator}
    \1 Direct and role-based allocation (although roles are not present in the test dataset)
    \1 Separation of concerns pattern and consequent allocation
    \1 Allocation based on seniority
    \1 Allocation based on risk class and loan goal \todo{Still to be impl.}
    \1 \todo[inline]{Check for completeness}
\end{outline}
\end{comment}

%\optnewpage

%\section{Discussion \& Conclusion}\label{sec:concl}

\subsection{Limitations and Future Work}

% \paragraph{Prototype}
While the prototype demonstrates the general feasibility and usefulness of the KRAFT framework, its complexity and scope are rather limited, yielding interesting challenges for future research. % \todo{Consider reframing this subsection in a bit of a less prototype focussed way}
First, on the knowledge input side, the realization of the conversational knowledge input interface as described in \cref{sec:input} can be researched. Further, integration with existing organizational mining techniques and the development of knowledge-graph-based ones can be investigated. 
Second, more sophisticated reasoning methods need to be inspected. 
Conceptually, approaches that look ahead, e.g., into the expected process continuation and future resource (un)availability, consider a broader range of allocation problems such as many-to-many allocation, or proactively learn, e.g., from user decisions and occurrences, can provide more optimal allocations. 
Technologically, integrating subsymbolic neural approaches is promising to achieve this, e.g., by utilizing remaining-time predictions as a decision input.   
Lastly, an investigation into run-time complexity needs to be launched, as decision-making during process execution needs close-to-real-time system reasoning. 

% \iodo{Imo, the conceptual level is more important here than the prototype; the extent of discussion should reflect that => Extend! \\ Insight: Many of the future work for the prototype is future work in general and might only need reframing}

Apart from examining how to implement the framework, further evaluation and validation are needed. 
We want to especially highlight the human factor here. As a major part of allocation knowledge currently exists in the form of implicit human knowledge, the interplay between these humans and systems w.r.t. to knowledge transfer, contradictory knowledge, but also impact on an emotional level needs to be further investigated\cite{klessascheck2024CriticalInvestigationRationalities}. 
Further, this paper only used the KRAFT framework in one reality-inspired use case. An evaluation with further use cases would help to strengthen the validity of the framework and potentially provide valuable pointers for improvement.

\section{Conclusion} \label{sec:concl}
%\todo{Maybe also restate a bit of the motivation there}
In this paper, we presented KRAFT, a framework for resource allocation based on knowledge graph reasoning. 
The framework provides guidance on how to integrate domain knowledge, process mining insights, and knowledge graph reasoning with business process management systems for resource allocation. 
Using a prototypical proof of concept, we demonstrated that such an approach has potential to support resource allocation in three major dimensions. 
First, to capture a wide range of allocation knowledge from various sources and integrate them into one data structure. 
Second, to easily adapt the stored knowledge on the relevant entities and mechanisms for allocation. 
Third, to get automated, comprehensive reasoning for selecting the best suitable resources for tasks.% \NM{Shouldn't we refer to explainability here (cf. the third objective)? \\LB: I guess it's fine considering the wording of O3} 
We argue that existing methods and technologies can serve as a basis for implementing the framework
and highlight interesting opportunities for further research. 

%\begin{outline}
%    \1 Argue that the basic building blocks for more mature implementations of the individual components are given
%    \1 Interesting opportunities for further research
%    \1 Other wording: To conclude, knowledge graph reasoning for resource allocation is feasible and useful for explainable reasoning on easily adaptable and holistic allocation knowledge. Sth. sth. future research
%\end{outline}

\optnewpage
\bibliographystyle{splncs04}
\bibliography{main}

\begin{thebibliography}{10}
\providecommand{\url}[1]{\texttt{#1}}
\providecommand{\urlprefix}{URL }
\providecommand{\doi}[1]{https://doi.org/#1}

\bibitem{annane2019BBOBPMN20}
Annane, A., {Aussenac-Gilles}, N., Kamel, M.: {{BBO}}: {{BPMN}} 2.0 {{Based
  Ontology}} for {{Business Process Representation}}. In: 20th European
  Conference on Knowledge Management ({{ECKM}} 2019). vol.~1, pp. 49--59 (2019)

\bibitem{arias2018TaxonomyHumanResource}
Arias, M., {Munoz-Gama}, J., Sep{\'u}lveda, M.: Towards a {{Taxonomy}} of
  {{Human Resource Allocation Criteria}}. In: Business {{Process Management
  Workshops}}, vol.~308, pp. 475--483. Springer, Cham (2018)

\bibitem{bachhofner2022AutomatedProcessKnowledge}
Bachhofner, S., Kiesling, E., Revoredo, K., Waibel, P., Polleres, A.: Automated
  {{Process Knowledge Graph Construction}} from {{BPMN Models}}. In: Database
  and {{Expert Systems Applications}}, vol. 13426, pp. 32--47. Springer, Cham
  (2022)

\bibitem{bein2025KnowledgeGraphsKey}
Bein, L., Pufahl, L.: Knowledge {{Graphs}}: {{A Key Technology}} for
  {{Explainable Knowledge-Aware Process Automation}}? In: Business {{Process
  Management Workshops}}, vol.~534, pp. 18--30. Springer, Cham (2025)

\bibitem{berti2023pm4py}
Berti, A., van Zelst, S., Schuster, D.: Pm4py: A process mining library for
  python. Software Impacts  \textbf{17},  100556 (2023)

\bibitem{cabanillas2015ralph}
Cabanillas, C., Knuplesch, D., Resinas, M., Reichert, M., Mendling, J.,
  Ruiz-Cort{\'e}s, A.: Ralph: a graphical notation for resource assignments in
  business processes. In: Advanced Information Systems Engineering: CAiSE 2015,
  Stockholm, Sweden, June 8-12, 2015, Proceedings 27. pp. 53--68. Springer
  (2015)

\bibitem{cabanillas2015SpecificationAutomatedDesigntime}
Cabanillas, C., Resinas, M., {del-R{\'i}o-Ortega}, A., {Ruiz-Cort{\'e}s}, A.:
  Specification and automated design-time analysis of the business process
  human resource perspective. Information Systems  \textbf{52},  55--82 (2015)

\bibitem{chen2022OverviewKnowledgeGraph}
Chen, Y., Li, H., Li, H., Liu, W., Wu, Y., Huang, Q., Wan, S.: An {{Overview}}
  of {{Knowledge Graph Reasoning}}: {{Key Technologies}} and {{Applications}}.
  Journal of Sensor and Actuator Networks  \textbf{11}(4), ~78 (2022)

\bibitem{corea2017DetectingComplianceBusiness}
Corea, C., Delfmann, P.: Detecting {{Compliance}} with {{Business Rules}} in
  {{Ontology-Based Process Modeling}}. Wirtschaftsinformatik 2017 Proceedings
  (2017)

\bibitem{corea2021OntologyBasedProcessModelling}
Corea, C., Fellmann, M., Delfmann, P.: Ontology-{{Based Process Modelling}} -
  {{Will We Live}} to {{See It}}? In: Conceptual {{Modeling}}. pp. 36--46.
  Springer, Cham (2021)

\bibitem{delong2024NeurosymbolicAIReasoning}
DeLong, L.N., Mir, R.F., Fleuriot, J.D.: Neurosymbolic {{AI}} for reasoning
  over knowledge graphs: A survey. IEEE Transactions on Neural Networks and
  Learning Systems pp. 1--21 (2024)

\bibitem{difrancescomarino2008ReasoningSemanticallyAnnotated}
Di~Francescomarino, C., Ghidini, C., Rospocher, M., Serafini, L., Tonella, P.:
  Reasoning on {{Semantically Annotated Processes}}. In: Service-{{Oriented
  Computing}} -- {{ICSOC}} 2007, vol.~4749, pp. 132--146. Springer Berlin
  Heidelberg (2008)

\bibitem{dumas2023AIAugmentedBusinessProcess}
Dumas, M., Fournier, F., Limonad, L., Marrella, A., Montali, M., Rehse, J.R.,
  Accorsi, R., Calvanese, D., De~Giacomo, G., Fahland, D., Gal, A., La~Rosa,
  M., V{\"o}lzer, H., Weber, I.: {{AI-Augmented Business Process Management
  Systems}}: {{A Research Manifesto}}. ACM Transactions on Management
  Information Systems  \textbf{14}(1),  1--19 (2023)

\bibitem{dumas2018FundamentalsBusinessProcess}
Dumas, M., La~Rosa, M., Mendling, J., Reijers, H.A.: Fundamentals of {{Business
  Process Management}}. Springer Berlin Heidelberg, Berlin, Heidelberg (2018)

\bibitem{havur2016ResourceAllocationDependencies}
Havur, G., Cabanillas, C., Mendling, J., Polleres, A.: Resource {{Allocation}}
  with {{Dependencies}} in {{Business Process Management Systems}}. In:
  Business {{Process Management Forum}}, vol.~260, pp. 3--19. Springer, Cham
  (2016)

\bibitem{hogan2022KnowledgeGraphs}
Hogan, A., Blomqvist, E., Cochez, M., {d'Amato}, C., {de Melo}, G., Gutierrez,
  C., Gayo, J.E.L., Kirrane, S., Neumaier, S., Polleres, A., Navigli, R.,
  Ngomo, A.C.N., Rashid, S.M., Rula, A., Schmelzeisen, L., Sequeda, J., Staab,
  S., Zimmermann, A.: Knowledge {{Graphs}}. ACM Computing Surveys
  \textbf{54}(4),  1--37 (2022)

\bibitem{ihde2022framework}
Ihde, S., Pufahl, L., V{\"o}lker, M., Goel, A., Weske, M.: A framework for
  modeling and executing task-specific resource allocations in business
  processes. Computing  \textbf{104}(11),  2405--2429 (2022)

\bibitem{klessascheck2024CriticalInvestigationRationalities}
Klessascheck, F., Bein, L., Haase, J., Pufahl, L.: A {{Critical Investigation}}
  of {{Rationalities}} in {{Automation}} with {{BPM}}. In: 2024 26th
  {{International Conference}} on {{Business Informatics}} ({{CBI}}). pp.
  30--39. IEEE (2024)

\bibitem{martin2024UnveilingUseCases}
Martin, N., Beerepoot, I.: Unveiling {{Use Cases}} for {{Human Resource
  Mining}}: {{A Framework}} of {{Past}} and {{Future Research}}. Business \&
  Information Systems Engineering  (2024)

\bibitem{ouyang2010ModellingComplexResource}
Ouyang, C., Wynn, M.T., Fidge, C.: Modelling {{Complex Resource Requirements}}
  in {{Business Process Management Systems}}. ACIS 2010 Proceedings  (2010)

\bibitem{pika2017MiningResourceProfiles}
Pika, A., Leyer, M., Wynn, M.T., Fidge, C.J., Hofstede, A.H.M.T., Aalst,
  W.M.P.V.D.: Mining {{Resource Profiles}} from {{Event Logs}}. ACM
  Transactions on Management Information Systems  \textbf{8}(1),  1--30 (2017)

\bibitem{pufahl2025automatic}
Pufahl, L., Stiehle, F., Ihde, S., Weske, M., Weber, I.: Resource allocation in
  business process executions---{{A}} systematic literature study. Information
  Systems p. 102541 (2025)

\bibitem{reichert2012enabling}
Reichert, M., Weber, B.: Enabling flexibility in process-aware information
  systems: challenges, methods, technologies, vol.~54. Springer (2012)

\bibitem{rubensson2025ConceptualFrameworkResource}
Rubensson, C., Pufahl, L., Mendling, J.: A {{Conceptual Framework}} for
  {{Resource Analysis}} in {{Process Mining}}. In: Enterprise {{Design}},
  {{Operations}}, and {{Computing}}. {{EDOC}} 2024 {{Workshops}}, vol.~537, pp.
  183--202. Springer, Cham (2025)

\bibitem{ruiz2021IdentificationDiseaseTreatment}
Ruiz, C., Zitnik, M., Leskovec, J.: Identification of disease treatment
  mechanisms through the multiscale interactome. Nature Communications
  \textbf{12}(1), ~1796 (2021)

\bibitem{russel2004workflow}
Russell, N., {Hofstede, ter}, A., Edmond, D., {Aalst, van der}, W.: Workflow
  Resource Patterns. {{BETA}} Publicatie : Working Papers, TU Eindhoven (2004)

\bibitem{tiddi2022KnowledgeGraphsTools}
Tiddi, I., Schlobach, S.: Knowledge graphs as tools for explainable machine
  learning: {{A}} survey. Artificial Intelligence  \textbf{302},  103627 (2022)

\bibitem{vandongen2017bpic}
{van Dongen}, B.: {{BPI}} challenge 2017 (2017)

\bibitem{verenich2019SurveyCrossbenchmarkComparison}
Verenich, I., Dumas, M., Rosa, M.L., Maggi, F.M., Teinemaa, I.: Survey and
  {{Cross-benchmark Comparison}} of {{Remaining Time Prediction Methods}} in
  {{Business Process Monitoring}}. ACM Transactions on Intelligent Systems and
  Technology  \textbf{10}(4),  1--34 (2019)

\bibitem{wang2019ExplainableReasoningKnowledge}
Wang, X., Wang, D., Xu, C., He, X., Cao, Y., Chua, T.S.: Explainable
  {{Reasoning}} over {{Knowledge Graphs}} for {{Recommendation}}. Proceedings
  of the AAAI Conference on Artificial Intelligence  \textbf{33}(01),
  5329--5336 (2019)

\bibitem{xu2013incorporating}
Xu, J., Liu, C., Zhao, X., Ding, Z.: Incorporating structural improvement into
  resource allocation for business process execution planning. Concurrency and
  Computation: Practice and Experience  \textbf{25}(3),  427--442 (2013)

\bibitem{zhu2024LLMsKnowledgeGraph}
Zhu, Y., Wang, X., Chen, J., Qiao, S., Ou, Y., Yao, Y., Deng, S., Chen, H.,
  Zhang, N.: {{LLMs}} for knowledge graph construction and reasoning: Recent
  capabilities and future opportunities. World Wide Web  \textbf{27}(5), ~58
  (2024)

\bibitem{zurmuehlen2004OrganizationalManagementWorkflow}
Zur~Muehlen, M.: Organizational {{Management}} in {{Workflow Applications}} --
  {{Issues}} and {{Perspectives}}. Information Technology and Management
  \textbf{5}(3/4),  271--291 (2004)

\end{thebibliography}

\begin{comment}

\optnewpage

\setcounter{tocdepth}{4}
\tableofcontents
\optnewpage
\setcounter{tocdepth}{1}
\listoftodos

\end{comment}
\end{document}